\documentclass[preprint,preprintnumbers,showpacs,showkeys,amsmath,amssymb,nofootinbib,superscriptaddress]{revtex4}

\usepackage{amsmath}
\def\[{\left\lbrack}
\def\]{\right\rbrack}

\def\({\left(}
\def\){\right)}
\def\ni{\noindent}

\newcommand{\be}{\begin{equation}}
\newcommand{\ee}{\end{equation}}
\newcommand{\ea}{\end{eqnarray}}
\newcommand{\ba}{\begin{eqnarray}}

\newcommand{\pa}{\partial}

\begin{document}

\title{Abelian and non-Abelian considerations upon compressible fluids with Maxwell-type equations and the minimal coupling with electromagnetic field}

%


\author{Everton M. C. Abreu}\email{evertonabreu@ufrrj.br}
\affiliation{Grupo de F\' isica Te\'orica e Matem\'atica F\' isica, Departamento de F\'{i}sica, Universidade Federal Rural do Rio de Janeiro, 23890-971, Serop\'edica, RJ, Brazil}
\affiliation{Departamento de F\'{i}sica, Universidade Federal de Juiz de Fora, 36036-330, Juiz de Fora, MG, Brazil}
\author{Jorge Ananias Neto}\email{jorge@fisica.ufjf.br}
\author{Albert C. R. Mendes}\email{albert@fisica.ufjf.br}
\affiliation{Departamento de F\'{i}sica, Universidade Federal de Juiz de Fora, 36036-330, Juiz de Fora, MG, Brazil}

\date{\today}

\vspace{3cm}

\begin{abstract}
\ni In this work we have obtained Maxwell-type equations for a compressible fluid which sources are functions of velocity and vorticity.  A correlation function and the dispersion relation were analyzed as function of the Reynolds number.  A Lagrangian for the Lamb vector and the vorticity was constructed and the equations of motion were discussed.  After that, we have analyzed the case of a charged fluid dynamics.  Finally, the non-Abelian generalization of some results was introduced.   A basic review for non-Abelian fluids was described in the Appendix.
\end{abstract}

\maketitle

\section{Introduction}

The Maxwell electrodynamics is the classical field theory that is the basics for the fluid's dynamics.  Since the Maxwell theory has passed through a non-Abelian generalization, the same is expected concerning fluid's dynamics.  The top of this non-Abelian evolution is the so-called quark-gluon plasma (QGP) which is obtained from high energy collisions of heavy nuclei.  This state of matter has been attracted big interest both theoretically and in experiments at the Relativistic Heavy Ion Collider (RHIC) facility and at CERN \cite{jackiw}.

It is well known that QGP liquid is dense.   However, apparently it flows with very little viscosity close to an ideal fluid which is ruled by the standard laws of hydrodynamics.   Hence, to analyze the results that comes from the RHIC one has to adopt a hydrodynamic point of view to the plasma for constructing its properties.   To consider the non-Abelian charges of the quarks and gluons is important in order to describe the dynamics for the proper fluid of the quarks and gluons.

We have in the literature several theories about some aspects of an ideal fluid in interaction with Yang-Mills fields \cite{heinz,choquet,hk,patten,bi}.   One of the objectives to create QGP from high energy heavy-ion collisions, is to test the prediction of a transition from a color-confining to an unconfining phase in QCD at high temperature and/or density.

Recent works have introduced an alternative way to describe fluid dynamics, compressible fluids \cite{Kambe} and the equations of a plasma \cite{Thompson}.   These results were obtained through the recasting of the equations and by obtaining a set of Maxwell-type equations for the fluid. 
This transformation in the structure of the equations implies the generalization of the idea of charge and current related to fluid dynamics \cite{Marmanis}. The understanding of what will be considered as a source term in the final formalism depends on the choice of the quantities and will be the main issue of this new structure of fluid dynamics. 

In Lighthill works about sound \cite{lighthill}, the applied stress tensor is considered as the source of the radiation field. 
The objective here is to construct, analogously to Marmanis  concerning an incompressible fluid, the Maxwell-type equations for compressible fluid considering the presence of a dissipation term from the beginning of the process. In this case we can see that there is an underlying difference between the definitions of the sources terms. 
Another important point that we will analyze, is the possibility of obtaining a Lagrangian description for the compressible fluid. 

In \cite{Thompson} the author has recently introduced an extension of this new structure of the plasma equations, for each kind of fluid, beginning with the equations of motion that  describe such system. We can obtain the equations by extending the Lagrangian description relative to the compressible fluid.   We will  consider a charged fluid and its coupling with the electromagnetic field.   We will combine both the Lamb vector and vorticity with the electric and magnetic fields.   As we will see, this coupling will result in some changes of the set of Maxwell-type equations that describe the system and also we will obtain new source terms.

The paper is organized in the following way: in section II we have obtained the Maxwell-type equations as functions of the vorticity and Lamb vector.  In section III we have computed the correlation function of the velocity in order to connect it with the current density.  In section IV we described the Lagrangian formalism for the theory above and we have constructed the non-Abelian version for our compressible fluid.   The analysis for a charged fluid was developed in section V.    The conclusions are depicted in section VII.

\section{Maxwell-type equations for a compressible fluid}

There are several works in the literature that show the similarity between fluid dynamics and classical electrodynamics. More recently this analogy has been used to write the equations of motion of a fluid with the same structure as Maxwell's equations of electromagnetism. Marmanis \cite{Marmanis} has described the dynamical behavior of average flow quantities in incompressible fluid flows with high Reynolds numbers. Kambe \cite{Kambe} has presented a generalization of a previous work concerning a compressible fluid.

Our purpose in this section is to obtain Maxwell-type equations considering the viscosity from the beginning and to construct a system of equations as functions of both the vorticity $(\vec{\rm {\omega}})$ and Lamb vector $(\vec {\rm l}=\vec{\rm {\omega}}\times \vec{\rm u})$ which is also known as vortex force.   In this way we will follow a different path from Kambe \cite{Kambe}

The equations of motion of a fluid are given by the Euler equation:

\be
\label{01}
{{\partial \vec{\rm u}}\over{\partial t}}+{\vec {\rm \omega}} \times {\vec{\rm u}}+\nabla \left(  {{1}\over{2}}\rm u^{2}\right) = -{{1}\over{\rho}}\nabla p
\ee

\ni where $1/2\,u^2$ is the kinetic energy, and this equation is supplemented by the following equations such as the continuity equation given by

\be
\label{02}
{{\partial \rho}\over{\partial t}} +\nabla (\vec{\rm u} \rho)=0,
\ee

\ni and the entropy equation,

\be
\label{03}
{{\partial s}\over{\partial t}} +\vec{\rm u}\cdot\nabla s =0,
\ee

\ni  where $\rho$ is the fluid density, $s$ is the entropy per unit mass, $p$ is the pressure, and $\vec {\rm \omega} =\nabla \times \vec{\rm u}$ is the vorticity. 

We can rewrite Eq.  (\ref{01}) by introducing the viscosity through the stress tensor 

\be\label{04}
\sigma_{ij}=\mu e_{ij} +\xi \delta_{ij}D,
\ee
\be\label{05}
e_{ij} =\partial_{j} {\rm u}_{i} +\partial_{i} {\rm u}_{j} -{{2}\over{3}} \delta_{ij} D \,\,\,\,\,  D\equiv \partial_{m} {\rm u}_{m},
\ee

\ni where the coefficients $\mu$ and $\xi$ are called coefficients of viscosity (and $\xi$ is also known as the second viscosity).  We have  from thermodynamics that 

\be\label{06}
-{{1}\over{\rho}} \nabla p = -\nabla h + T\nabla s,
\ee

\ni where $h$ is the enthalpy per unit mass and $T$ is the temperature. So, we obtain

\be\label{07}
{{\partial \vec{\rm u}}\over{\partial t}} + \vec{\rm {\omega}}\times \vec{\rm u} +\nabla\left( {{1}\over{2}} {\rm u}^2 \right) = -\nabla h + \vec{\kappa},
\ee

\ni where

\be\label{08}
\vec{\kappa} = T\nabla s +{{1}\over{\rho}}\nabla \sigma
\ee

\ni and Eq. (\ref{07}) is the Navier-Stokes equation for a viscous compressible fluid \cite{Landau}, different from Eq. (\ref{01}).   From a dimensional analysis of Eq. (\ref{08}) we see that $\vec{k}$ has acceleration dimension.

Hence, we will define from Eq. (\ref{01}) the Lamb vector as
\be\label{09}
\vec {\rm l} = -{{\partial \vec{{\rm u}}}\over{\partial t}} - \nabla \Omega + \vec{\kappa},
\ee
where 
\be\label{10}
\Omega = h + {{1}\over{2}}{\rm u}^2
\ee
is the total energy.   From Eq. (\ref{09}) we see that the sign of $\nabla \cdot \vec{l}$ oscillates between negative and positive, which occurs usually in turbulent flows.

Now, let us compute the divergence, curl, and time derivative of the Lamb vector and after that, the divergence of the vorticity to obtain respectively that

\be\label{11}
\nabla \cdot \vec {\rm l} ={n} + \nabla \cdot \vec{\kappa},
\ee
\be\label{12}
\nabla \times \vec {\rm l} +{{\partial\vec{\rm {\omega}}}\over{\partial t}} = \nabla \times \vec{\kappa},
\ee
\be\label{13}
{{\partial \vec {\rm l}}\over{\partial t}} -{\rm u}^2 \nabla \times \vec{\rm {\omega}} = -\, \vec{j} + {{\partial \vec{\kappa}}\over{\partial t}}
\ee
\be\label{14}
\nabla \cdot \vec{\rm {\omega}} = 0,
\ee

\ni where $\vec{j}$ will be defined in a moment and

\be
\label{15}
n\,=\,-\,\frac{\pa}{\pa t}  \nabla \cdot \vec{\rm u} \,-\,\nabla^2 \Omega\,\,
\ee

\ni is connected with the concept of {\it source terms}, analogous to both the electric charge and current densities.   Notice that $n$ is directly proportional to the time variation of the dilatation and the spatial variation of the energy.   On the other hand, it is well known that  $$\nabla \cdot \vec{l} \,=\, \vec{u} \cdot \nabla \times \vec{\omega}\,-\,\vec{\omega} \cdot \vec{\omega}$$ and from (\ref{11}) we can write that $$n\,=\,\vec{u} \cdot \nabla \times \vec{\omega}\,-\,\vec{\omega} \cdot \vec{\omega}$$ when $\nabla \cdot \vec{k}$ and it means that the source term can be given can be given by the velocity and vorticity.

From the Lagrange acceleration formula \cite{truesdell} given by  $$ \dot{\vec{u}}\,=\,\frac{\pa \vec{u}}{\pa t}\,+\,\vec{\omega} \times \vec{u}\,+\,\nabla \Big( \frac 12 u^2 \Big)\quad \Longrightarrow\quad \nabla \times \dot{\vec{u}}\,=\,\frac{\pa \vec{\omega}}{\pa t}\,+\, \nabla \times \vec{l}\,\,,$$ which is analogous to Eq. (12), and as we have said above, $\,\vec{k}$ is the acceleration.

In Eq. (\ref{11}) we have that the sources of the Lamb vector are the pressure, enthalpy and velocity gradients. In Eq. (\ref{12}) one can show that the time variation of the angular velocity is equal to the torque given by the Coriolis force which is the Lamb vector. 
The last equation represents the conservation of the vorticity flux along a tube of vorticity. 
Analogously to the electromagnetic case, we can say that it means the absence of monopole source of vorticity.

Moreover, back to Eq. (\ref{12}), in the case of a motion with steady vorticity ($\pa \omega / \pa t = 0$) and a lamellar Lamb vector, i.e., $\nabla \times \vec{l} = 0$, we have, using Eq. (\ref{08}), that $$\nabla T \times \nabla s \,=\,-\,\nabla \frac{1}{\rho} \times \nabla \sigma\,\,,$$ which is the condition for the second and third vorticity theorems of Helmholtz \cite{truesdell}.   The new thing is that this last equation is the condition for circulation-preserving motion independent of $\vec{\omega}$ and $\vec{l}$.

Having said that, the Lamb  vector and the vorticity should be taken as the kernel of turbulent dynamics rather through velocity and vorticity fields or velocity and pressure fields \cite{Marmanis}. Then any term that cannot be explicitly expressed as a function of only $\vec{\rm w}$ or $\vec{\rm l}$, will be treated as a source term. Within the term connected with the viscosity through the {\it source term} in Eq. (\ref{15}) and



\be\label{16}
\vec{j} = \vec{{\rm u}}{n} + \nabla \times (\vec{{\rm u}} \cdot \vec{\rm {\omega}})\vec{{\rm u}} + \vec{\rm {\omega}}\times\nabla \Big(\Omega +{\rm u}^2 \Big)
+ 2\, \Big[(\vec{\rm {\omega}}\times \vec{{\rm u}}) \cdot \nabla \Big] \vec{{\rm u}} -(\vec{\rm {\omega}}\times \vec{{\rm u}})(\nabla \cdot \vec{{\rm u}})\,\,,
\ee

\ni where we have to notice that the source terms in Eqs. (\ref{15}) and (\ref{16}), differ from the source terms given in \cite{Marmanis} thanks to the presence of $\nabla \cdot \vec{{\rm u}}$,  since the fluid is compressible. The charge density $n$, related to the vorticity can be regarded as a topological feature of the flow \cite{Curtis}.


An important observation can be made about the presence of the term $\vec{\kappa}$ in Eq. (\ref{09}) and its effect into Maxwell-type equations, which can be rewritten as
\be\label{17}
\nabla \cdot \left(\vec {\rm l}-\vec{\kappa}\right) ={n},
\ee
\be\label{18}
\nabla \times \left(\vec {\rm l}-\vec{\kappa}\right) +{{\partial\vec{\rm {\omega}}}\over{\partial t}} = 0,
\ee
\be\label{19}
{{\partial (\vec {\rm l}-\vec{\kappa})}\over{\partial t}} -{\rm u}^2 \nabla \times \vec{\rm {\omega}} = - \vec{j} 
\ee
\be\label{20}
\nabla \cdot \vec{\rm {\omega}} = 0.
\ee
We can see that the presence of this $\vec{\kappa}$-term, in analogy with the Maxwell electromagnetism, shows that $\vec\kappa $ acts as a polarization vector for the Lamb vector, such as the $\vec P$ vector for the electric field in classical electrodynamics. So, let us construct a new Lamb vector field $\vec{{\rm l}}^{\prime}=\vec {\rm l} -\vec{\kappa}$ in equations (\ref{17})-(\ref{20}) and consequently we re-obtain the same set in (\ref{11})-(\ref{14}) for the vectors $\vec{{\rm l}}^{\prime}$ e $\vec {\rm w}$  \cite{Kambe}. 

The system of equations (\ref{17})-(\ref{20}) share a common feature with the so-called microscopic Maxwell equations concerning the extremely fast space and time variable fields . As in \cite{Marmanis} for an incompressible fluid, we will use the spatial filtering method proposed by Russakoff \cite{Russakoff} to obtain the dynamical behavior of average flow quantities. 
After the application of  the process, we have that

\be\label{21}
\nabla \cdot \langle\vec{{\rm l}}^{\prime}\rangle =\langle{n}\rangle ,
\ee
\be\label{22}
\nabla \times \langle\vec{{\rm l}}^{\prime}\rangle +{{\partial\langle\vec{\rm {\omega}}\rangle}\over{\partial t}} = 0,
\ee
\be\label{23}
{{\partial \langle\vec{{\rm l}}^{\prime}\rangle}\over{\partial t}} -\langle{{\rm u}^2}\rangle \nabla \times \langle\vec{\rm {\omega}}\rangle = - \langle\vec{j}\rangle 
\ee
\be\label{24}
\nabla \cdot \langle\vec{\rm {\omega}}\rangle = 0,
\ee

\ni where $\langle {\rm u}^2 \rangle =a^2$ is the spatial averaged squared velocity. 

We can promote an extension of this analogy by introducing the potentials which, in this case, correspond to the average velocity field $\langle \vec {\rm u}\rangle$ (vector potential) and energy function $\langle \Omega\rangle$ (the scalar potential). As in Maxwell's electromagnetism, sometimes it is more appropriate to work with the equations involving only the potential in order to keep a smaller number of second-order equations, rather than against a set of coupled first-order partial differential equations (\ref{21})-(\ref{24}). So,  from Eq. (\ref{23}) and using the expression in (\ref{09}), it can be shown that the average velocity field obeys the following equation

\be\label{24.3}
\nabla^2 \langle \vec {\rm u} \rangle - {{1}\over{a^2}}{{\partial^2 \langle \vec {\rm u}\rangle}\over{\partial t^2}} + \nabla \left( \nabla \cdot \langle \vec {\rm u}\rangle + {{1}\over{a^2}}{{\partial \langle \Omega \rangle}\over{\partial t}} \right)= -a^{-2}\langle \vec j \rangle, 
\ee  

\ni concerning the average velocity, and

\be\label{24.4}
\nabla^2 \langle \Omega \rangle + {{\partial}\over{\partial t}}\nabla \cdot \langle \vec {\rm u} \rangle = -\,\langle \hat n \rangle,
\ee

\ni for the energy function $\langle \Omega \rangle$ (see Eq. \ref{15}).

The Eq. (\ref{24.3}) shows the wave character of the fluid dynamics equations described by Eqs. (\ref{21})-(\ref{24}).  Besides, Eqs. (\ref{24.3}) and (\ref{24.4}) allow us to make an interesting observation about a conceptual difference between both theories, namely, electromagnetism and fluid theories.

The equations for the electromagnetic potentials  $(\vec A, \Phi)$, which have expressions analogous to mathematical equations, can be decoupled through a suitable choice for potential called gauge transformations

\be
\label{24.5}
\nabla \cdot \vec A =0 \,\,\,\,\, ({\rm Coulomb\,\, gauge})
\ee
and
\be\label{24.6}
\nabla \cdot \vec A + {{1}\over{c^2}} {{\partial \Phi}\over{\partial t}} = 0 \,\,\,\,\,\, ({\rm Lorentz\,\, gauge}),
\ee

\ni which do not affect the physics of the system.
In fluid dynamics, this freedom is not simply a choice of gauge, it has implications about the physical nature of the flow. 
The relation $\nabla \cdot \langle \vec {\rm u}\rangle=0$  in fluid dynamics, which is equivalent to the Coulomb gauge, is not a true gauge but it is a choice of the incompressibility of the flow.

Similarly, the Lorentz gauge has a corresponding equation which connects the fluid dynamics given by 

\be\label{24.7}
\nabla \cdot \langle \vec{\rm u} \rangle + {{1}\over{a^2}}{{\partial \langle \Omega \rangle}\over{\partial t}} =0,
\ee

\ni and which is relative to a compressible fluid. Thus, we observe that a ``gauge choice" in fluid dynamics is directly related to the hypothesis made about the compressibility or incompressibility of the fluid \cite{Thompson}.

A mathematical object of great importance in statistical theory of fluid dynamics is the correlation function of two points of the velocity field defined as

\be\label{24.8}
R_{ij}(\vec r)\equiv \langle v_i (\vec x)v_j (\vec x + \vec r) \rangle
\ee

\ni for two points separated by the displacement vector $\vec r$.  Both assumptions of stationary and homogeneity allow us to establish, as indeed has already been attempted in the link above, this correlator does not depend on time, but only on the displacement vector $\vec r$. The space-time correlation are underlying in fluid dynamics and has a wide application \cite{Xin}.

\section{The correlation function}

Great efforts have been devoted during decades to the discovery of how the kinetic energy of a turbulent flow is partitioned along the length scales. The focus of these studies is the energy spectrum of the turbulent flow in Fourier space, obtained directly from the correlation function for speed.

Our purpose now is to show that we can connect the correlation function of the velocity field with the current density $\vec j$.   To accomplish that we have to consider a specific case which is not simpler than the flow of a compressible fluid.   It is an incompressible fluid with high Reynolds numbers in the completely developed turbulent regime \cite{Frish}. 

In this case, the wave equation in (\ref{24.3}) (without the electromagnetic field coupling) can be whiten as

\be\label{24.9}
{{\partial^2 \vec v}\over{\partial t^2}} \,=\,a^2\nabla^2 \vec{v} \,+\, \vec j \,-\,\nabla \left( {{\partial \phi}\over{\partial t}} \right) \,+\,\nu^2 \nabla^4 \vec{v}\,\,, 
\ee 

\ni where $\nu$ is the (kinematic) viscosity, $\vec v =\langle \vec {\rm u} \rangle$ and $\phi =\langle \Omega \rangle$.

As in the electromagnetic theory, we can decompose $\vec j$ into its longitudinal part (irrotational) $\vec j_l$ and its transverse part (solenoidal) $\vec j_t$, i.e.,

\be\label{24.10}
\vec j = \vec{j}_l +\vec{j}_t\,\,,
\ee

\ni where we can show that

\be\label{24.11}
\vec{j}_l =\nabla {{\partial \phi}\over{\partial t}} 
\ee

\ni and we can rewrite Eq. (\ref{24.9}) as follows

\be\label{24.12}
{{\partial^2 \vec v}\over{\partial t^2}} =a^2\nabla^2 \vec{v}  +\nu^2 \nabla^4 \vec{v} +\vec{j}_t, 
\ee 

\ni which is a wave equation with a forcing term exclusively expressed in terms of the turbulence sources.

The non-homogeneous wave Eq. (\ref{24.12}) has a solution of the type 

\be\label{24.13}
\vec v(\vec x, t) = \int d^4{\bf x}\,G(\vec{x}, t\,;\,\vec{x}^{\prime},t) \vec{j}_t (\vec{x}^{\prime}, t^{\prime} )
\ee

\ni where the Green's function  in Eq. (\ref{24.13}) satisfies the inhomogeneous equation

\be\label{24.14}
\left( a^2 \nabla^2 + \nu^2 \nabla^4 -{{\partial^2}\over{\partial t^2}} \right) G(\vec{x}, t\,;\,\vec{x}^{\prime},t)= \delta({\vec x}- {\vec x}^{\prime})\delta(t-t^{\prime}).
\ee

The wave equation in (\ref{24.12}) describes waves with the following dispersion relation \cite{Marmanis},
\be\label{24.15}
\omega = a k\left(1 -{{\nu^2}\over{2a^2}} k^2 \right),
\ee

\ni where, in the upper limit for the value  of $k$ ($k_f$: filter microscale) we have

\begin{equation*}
{{\nu^2 k^2_f}\over{a^2}} =O(Re^{-1}).\nonumber
\end{equation*}

So, when we consider a fluid with high Reynolds number, we have a linear law to the dispersion relation: $\omega(k) = ak$. In this case, the operator in Eq. (\ref{24.14}) is the type D'Alembertian with the usual Green's function, such as

\be\label{24.16}
G(\vec{x}, t\,;\,\vec{x}^{\prime},t)=\left\{
\begin{array}{ccc}
{{1}\over{4\pi a^2}}{{\delta\left(t-t^{\prime} - {{|\vec{x} -\vec{x}^{\prime}|}\over{a}}\right)}\over{|\vec{x} -\vec{x}^{\prime}|}} &{\rm if}& (t>t^{\prime})\\
\\
0 &{\rm if}& (t<t^{\prime}).\\
\end{array}
\right.
\ee

Now, substituting Eq. (\ref{24.16}) into (\ref{24.13}) and solving the integration in $t^{\prime}$ we have that

\be\label{24.17}
\vec v(\vec x, t) = \int d^3{\vec x}^{\prime}\, {{{\vec j}_{t}\left(\vec{x}^{\prime},t - {{|\vec{x} -\vec{x}^{\prime}|}\over{a}}\right)}\over{|\vec{x} -\vec{x}^{\prime}|}},
\ee

\ni and with this result, we can write the correlation function as

\ba\label{24.18}
R_{ij}(\vec r)=\langle v_{i}(\vec x) v_{j}(\vec{x} +\vec r)\rangle &=& \int_{\vec x}^{\vec x +\vec r} d^{3} \vec{y} \int_{{\vec x}}^{{\vec x} + \vec r} d^3 \vec{z} \langle f_{i}(\vec y)f_j(\vec z)\rangle\nonumber \\
&=& \int_{\vec x}^{\vec x +\vec r} d^{3} \vec{y} \int_{{\vec x}-\vec{y}}^{{\vec x} + \vec r -\vec{y}} d^3 \vec{\rho} \,\langle f_{i}(\vec y)f_j(\vec y +\vec\rho)\rangle ,
\ea
where we have defined
\be\label{24.19}
\vec{f}({\vec x}^{\prime}; t)={{{\vec j}_{t}\left(\vec{x}^{\prime},t - {{|\vec{x} -\vec{x}^{\prime}|}\over{a}}\right)}\over{|\vec{x} -\vec{x}^{\prime}|}}.
\ee

We can observe in the Eq. (\ref{24.18}) that the correlation function of the velocity field depends on both the correlation function of the current density $\vec{j}_{t}$ and the displacement vector $\vec r$. The current density is an input and can not be determined by theory, it depends on both the observation and the geometry of the flow.

\section{Lagrangian formalism and minimal coupling}

From the geometric and Lie algebraic points of view, Arnold \cite{Arnold} has showed that the Euler flow can be described via the Hamiltonian formalism in any dimensions. This has a lot of interesting consequences concerning fluid mechanics \cite{arti18e19dobrazilian}. One of them is that it is the standard path to introduce statistical approaches into dynamical systems.   The Hamiltonian is helpful in order to build a statistical measure.
However, it is not quite obvious that this process can be used when viscosity is taken into account. 

In paper \cite{Albert1} one of us has shown that this is possible if we consider the equations of metafluid dynamics obtained by Marmanis \cite{Marmanis}. Since the structure of Eqs. (\ref{21})-(\ref{24}) are the same as the equations of metafluid dynamics, although we are working with a compressible fluid, we can similarly write down the Lagrangian density of the theory of a compressible fluid.   Hence, let us begin with

\be\label{25}
{\cal L} ={{1}\over{2}}\left( {\vec{l}}^2 -a^2 {\vec \omega}^2 \right) 
\ee

\ni where we will define the averaged Lamb vector $\vec{l} = \langle \vec {\rm l} \rangle$, as

\be\label{26}
\vec l = -{{\partial \vec {v}}\over{\partial t}} -\nabla \phi+ \vec{k}\,\,,
\ee

\ni where $\vec \omega=\langle \vec {\rm w} \rangle$ and $\vec{k}=\langle \vec \kappa \rangle$.  We can see at first sight that the Lagrangian in (\ref{25}) has kind of duality since we can substitute $l \rightarrow ia\omega$ and 
$a\omega \rightarrow il$ and the Lagrangian in (\ref{25}) is invariant under this symmetry.  However, one can ask about the physical meaning of the complex terms in the fluid dynamics.

So, we can also write the above Lagrangian density in terms of the potential in the following way

\be\label{27}
{\cal L} ={{1}\over{2}} \left(-{{\partial \vec v}\over{\partial t}} -\nabla \phi + \vec{k}\right)^2 -{{1}\over{2}}a^2 \left(\nabla \times \vec v\right)^2\,\,,
\ee

\ni which is the Lagrangian of a compressible fluid with nonzero viscosity, which gives us the Navier-Stokes equation directly from the conjugated momenta of the velocity field

\be\label{27.1}
\vec \pi = {{\delta \cal{L}}\over{\delta \dot{\vec v}}} = {{\partial \vec v}\over{\partial t}} +\nabla \phi -\vec{k}= -\vec l,
\ee

\ni when it is compared with the equation (\ref{26}).   Thus, we have obtained the Navier-Stokes equation for the mean field

\be\label{28}
{{\partial \vec v}\over{\partial t}} = \vec l -\nabla \phi -\vec {k},
\ee

\ni where $\vec l =\vec \omega \times \vec v$. Notice that the contributions of viscosity and $T\nabla s$ are embedded in $\vec k$. 

It is easy to see that the Lagrangian density (\ref{27}) gives us the set of Maxwell-type (\ref{11})-(\ref{14}) equations  for the homogeneous case (no source). Computing the Euler-Lagrange equations relative to $\phi$ and $\vec v$ from Eq. (\ref{27.1}), respectively, we can write that

\be\label{29}
\nabla \cdot ( -\nabla \phi -\dot{\vec v} +\vec {k} )=0,
\ee

\ni and using Eq. (\ref{26}), we can see that

\be\label{30}
\nabla \cdot \vec l =0.
\ee

For the velocity field $\vec v$, we have

\be\label{31}
{{\partial}\over{\partial t}}(-\nabla \phi -\dot{\vec v} +\vec {k}) =a^2 (\nabla \times \nabla \times \vec v)\,\,,
\ee

\ni and from Eq. (\ref{26}) and using that $\vec{\omega} = \nabla \times \vec v$, this last equation can be written as

\be\label{32}
{{\partial \vec l}\over{\partial t}}=a^2 (\nabla \times \vec\omega) +{{\partial \vec k}\over{\partial t}}.
\ee

The other two equations, (\ref{12}) and (\ref{14}), can be obtained directly from the $\vec \omega$ and $\vec l$ definitions. Taking the divergence of $\vec \omega$, we can obtain Eq. (\ref{14}).   To obtain Eq. (\ref{12}), we have to take the curl of Eq. (\ref{09}).

Now, let us consider the case where the sources do not vanish. In this case, the sources appear into the equations of motions and, as a consequence, we can write its Lagrangian density as

\be\label{33}
{\cal L} ={{1}\over{2}} \left(-{{\partial \vec v}\over{\partial t}} -\nabla \phi + \vec{k}\right)^2 -{{1}\over{2}}a^2 \left(\nabla \times \vec v\right)^2 
+ \vec{J} \cdot \vec{v} - N\phi
\ee

\ni where $N=\langle {n}\rangle$, $\vec J =\langle \vec j \rangle$ and $n$ was given in Eq. (\ref{17}).

Considering the term $\vec k$ as a vector polarization for the Lamb vector, let us define

\be\label{34}
{\vec{l}}^{\prime} =\langle \vec l -\vec \kappa \rangle= -{{\partial \vec v}\over{\partial t}} -\nabla \phi, 
\ee

\ni we can rewrite the Lagrangian density (\ref{27}) as

\be\label{35}
{\cal L} ={{1}\over{2}}\left( {\vec{l^{\prime}}}^2 -a^2 {\vec \omega}^2 \right) ={{1}\over{2}} \left(-{{\partial \vec v}\over{\partial t}} -\nabla \phi \right)^2 -{{1}\over{2}}a^2 \left(\nabla \times \vec v\right)^2\,\,.
\ee 

Therefore, if we look at the Lamb vector defined in (\ref{34}) and the mean vorticity 

\be\label{36}
\vec \omega =\nabla \times \vec v
\ee

\ni we can see that $\vec{l}^{\prime}$ and $\vec \omega$, as well as the electric and magnetic fields, are components of a second-rank tensor, the strength tensor, defined by

\be\label{37}
T^{\mu\nu}={\tilde\partial}^{\mu} U^{\nu} -{\tilde\partial}^{\nu}U^{\mu},
\ee

\ni where $\tilde{\partial}^{\mu}=\left( {{1}\over{a}}{{\partial}\over{\partial t}},\nabla \right)$ and the four-vector potential is

\be\label{38}
U^{\mu} \equiv (\phi,\,\, a\vec v).
\ee

\ni The nonzero components of the $T^{\mu\nu}$ are

\be\label{38.1}
T^{0i} ={l^{\prime}}^{i} \,\,\qquad{\rm and}\,\qquad T^{ij} =a\,\omega^{k}\,\,\,\quad {\rm {w}ith}\,\,\, i,j,k \,\, {\rm cyclic}.
\ee

\ni Thus we can see that, compared with the second-rank tensor in (\ref{37}), the second rank tensor introduced by Mahajan on \cite{Mahajan} is a relativistic extension of the Lamb vector and vorticity given in (\ref{38.1}).

Therefore, one can write (\ref{35}) using $T^{\mu\nu}$ as

\be\label{39}
{\cal L} =-{{1}\over{4}}T^{\mu\nu}T_{\mu\nu},
\ee

\ni or, considering the presence of the source terms, we have that

\be\label{40}
{\cal L} =-{{1}\over{4}}T^{\mu\nu}T_{\mu\nu} -{{1}\over{a}} J_{\mu}U^{\mu},
\ee

\ni where the four-vector $J^{\mu}$ is defined as

\be\label{41}
J^{\mu}\equiv (\,N,\,\,a\vec{J}\,)\,\,.
\ee

The inhomogeneous equations of the compressible fluid (\ref{21}) and (\ref{23}), in terms of $T^{\mu\nu}$ and the four-current $J^{\mu}$, can be connected through a covariant expression

\be\label{42}
\partial_\mu T^{\mu\nu} = {{1}\over{a}} J^{\nu}\,\,.
\ee

\ni Similarly, the homogeneous Eqs. (\ref{22}) and (\ref{24}) can be written in terms of the dual field-strength tensor as 

\be\label{43}
\partial_{\mu} {\cal T}^{\mu\nu}= 0
\ee

\ni where ${\cal T}^{\mu\nu} ={{1}\over{2}}\epsilon^{\mu\nu\alpha\beta}T_{\alpha\beta}$.

\subsection{The non-Abelian generalization}

In \cite{bmm} the authors carried out a NA generalization of \cite{Mahajan} in order to apply the formalism to QGP.   With the same motivation we will provide the NA version of these last results.  In the name of self-containment, we have provided the interested reader with the basic knowledge of the issue in the Appendix.  From now on we will assume that the reader have read the Appendix and have this basic knowledge.

Having said that, since $Q_a$ is the classical color charge of the particle, let us define the Lie algebra which is defined with anti-Hermitian Lie algebra objects with basis $T^a$
\ba
[T^a , T^b ]\,&=&\,f^{abc}\,T^c \nonumber \\
\mbox{tr}\,(T^a\, T^b ) \,&=&\,-\frac 12 \delta^{ab}
\ea

\ni which allows us to define the fluid velocity as
\ba
\vec{u}\,=\,\vec{u}^a\,T^a
\ea

\ni the vorticity as $\vec{\omega}=\vec{\omega}^a\,T^a$ where
\ba
\vec{\omega}^a\,=\,\nabla\,\times\,\vec{\omega}^a
\ea

\ni and the Lamb vector is $\vec{l}=\vec{l}^a\,T^a$ where
\ba
\vec{l}^a\,=\,\epsilon^{abc}\vec{\omega}^b \times \vec{u}^c\,\,.
\ea

\ni Hence we can rewrite the Lagrangian in Eq. (\ref{35}) as 
\ba
{\cal L}\,=\, \frac 12 (\vec{l}^{'2}_a\,-\,a^2 \vec{\omega}^2_a ) \,=\, \frac 12 \Big( -\frac{\pa \vec{v}_a}{\pa t}\,-\,\nabla \phi_a \Big)^2\,-\, \frac 12 a^2 (\nabla \times \vec{v}_a )^2
\ea

\ni where $\phi_a = Q_a <\Omega>$ and $\vec{v}_a = Q_a <\vec{u}>$.  The strength tensor in Eq. (\ref{37}) can be written in a NA way as
\ba
T^a_{\mu\nu}\,=\, \tilde{\pa}_{\mu} U^a_{\nu} \,-\, \tilde{\pa}_{\nu} U^a_{\mu} \,-\, i\,g [U_{\mu} ,\,U_{\nu} ]^a
\ea

\ni where $[U_{\mu},\,U_{\nu}]^a\,=\,f^a_{bc} U^b_{\mu} U^c_{\nu}$ and $g$ is the gauge charge.   The NA four-vector potential is given by
\ba
U^{\mu}_a\,=\,(\phi_a , a\vec{v}_a )
\ea

\ni where $T^{0ia}\,=\,l^{'ia}$ and $T^{ija}\,=\,a\,\omega^{ka}\:, \mbox{where}\:i,\,j,\,k\:=\mbox{cyclic}$, and the Lagrangian for the NA filed strength in Eq. (\ref{39}) 
\ba
{\cal L}\,=\,-\frac 14 T^{\mu\nu a} T^a_{\mu\nu}
\ea

\ni and, considering source terms we have that
\ba
{\cal L}\,=\,-\frac 14 T^{\mu\nu a} T^a_{\mu\nu}\,-\,\frac 1a J_{\mu}^a U^{\mu a}
\ea

\ni where $J^{\mu}_a\,=\,(N_a ,\,a \vec{J}_a )\,;\:N_a = Q_a <u>$ and $J^{\mu}_a \,=\,Q_a <j^{\mu}_a >$.   The non-Abelianization of Eqs. (\ref{42}) and (\ref{43}) is direct.   In the same manner we can promote the non-Abelianization of the elements when we consider a charged fluid as in section IV.

Since in sections IV and V we have used a gauge theoretical analogy \cite{bmm} to construct an electromagnetic-type theory as functions of the Lamb vector and vorticity, in this section we have developed a NA generalization to obtain a NA fluid-field system analogous to the well known Yang-Mills fluids.

\section{The electromagnetic interaction analysis}

Let us now consider the case where the fluid described by Eqs. (\ref{01})-(\ref{03}) is a charged fluid with each species labeled by the index $\alpha $. Besides, they can be described by a set of linear Maxwell-type equations for the mean field, such as 

\be\label{44}
\nabla \cdot {\vec{l^{\prime}}}_{\alpha} =N_{\alpha},
\ee

\be\label{45}
\nabla \times {\vec{l^{\prime}}}_{\alpha} +{{\partial\vec{\omega}_{\alpha}}\over{\partial t}} = 0,
\ee

\be\label{46}
{{\partial {\vec{l^{\prime}}}_{\alpha}}\over{\partial t}} -a^2 \nabla \times \vec{\omega}_{\alpha} = - \vec{J}_{\alpha} 
\ee

\be\label{47}
\nabla \cdot \vec{\omega}_{\alpha} = 0,
\ee

\ni where ${\vec{l^{\prime}}}_{\alpha} =\langle \vec {\rm l}_{\alpha} -\vec{\kappa}_{\alpha}\rangle$ and analogously to Eqs. (\ref{21})-(\ref{24}), which can be obtained from the Lagrangian density for each species given by

\be\label{48}
{\cal L} ={{1}\over{2}}\left( {\vec{l^{\prime}}}^{\,2}_{\alpha} -a_{\alpha}^2 \,{\vec \omega}_{\alpha}^{\,2} \right),
\ee

\ni where $a_{\alpha}^{2}=\langle \vec{\rm u}_{\alpha}^{2}\rangle$.

In recent works \cite{Mahajan,Thompson} the authors  present an extension of the analogy between fluid dynamics and classical electrodynamics by considering one multifluid plasma using different approaches.   However, in either case we have as the starting point the equations of motion. Our proposal is to consider the fluid immersed in an electromagnetic field and the interaction between them from the Lagrangian density in (\ref{48}). 

Let us then consider the following prescription for both the Lamb field $(\vec{l}_{\alpha})$ and the vorticity field $(\vec{\omega}_{\alpha})$ by defining the coupling with electromagnetic field such as

\be\label{49}
{\vec{l}}\,^{\prime}_{\alpha} \Longrightarrow \hat{\vec{l}}_{\alpha} = {\vec{l}}\,^{\prime}_{\alpha} + g\vec E
\ee

\ni and

\be\label{50}
\vec{\omega}_{\alpha} \Longrightarrow \hat{\vec{\omega}}_{\alpha}= \vec{\omega}_{\alpha}+ b\vec B\,\,,
\ee

\ni where $g$ and $b$ are the coupling constants that will be calculated later. 

Using Eqs. (\ref{49}) and (\ref{50}), we can now write the new Lagrangian density $\cal{L}^{\prime}$

\ba\label{51}
\cal{L}^{\prime}&=&{{1}\over{2}}\left(  \hat{\vec{l}}\,_{\alpha}^{2} -a_{\alpha}^2 \hat{\vec{\omega}}^{2}_{\alpha} \right)\nonumber \\
&=& {{1}\over{2}}\left(  {\vec{l^{\prime}}}_{\alpha} +g\vec E \right)^2 -{{1}\over{2}}a_{\alpha}^2 \left({\vec{\omega}}_{\alpha}+b\vec B \right)^{2} \nonumber\\
&=&{{1}\over{2}}\left(-{{\partial \vec{v}_{\alpha}}\over{\partial t}}-\nabla \phi_{\alpha} + \vec{k}_{\alpha} + g\vec E \right)^2 -{{1}\over{2}}a_{\alpha}^2 \left(\nabla \times \vec{v}_{\alpha} + b \vec B \right)^2,
\ea

\ni which is the Lagrangian density of a charged fluid immersed in an electromagnetic field where the term $\vec{k}_{\alpha}$ carries the contributions due to  viscosity  and statistical features $(T\nabla s)$ of the fluid. The conjugated momenta associated with velocity is computed as

\be\label{52}
\vec{\pi^{\prime}}_{\alpha} ={{\delta \cal{L}^{\prime}}\over{\delta \dot{\vec{v}}_{\alpha}}} = {{\partial \vec{v}_{\alpha}}\over{\partial t}}+\nabla \phi_{\alpha} - \vec{k}_{\alpha} - g\vec E = - \hat{\vec{l}}_{\alpha}
\ee

\ni or

\be\label{53}
{{\partial \vec{v}_{\alpha}}\over{\partial t}}+\nabla \phi_{\alpha} - \vec{k}_{\alpha} - g\vec E = - \hat{\vec{\omega}}_{\alpha} \times \vec{v}_{\alpha}.
\ee

\ni Using the relation (\ref{50}), we finally have that

\be\label{54}
{{\partial \vec{v}_{\alpha}}\over{\partial t}}+  \vec{\omega}_{\alpha} \times \vec{v}_{\alpha} = -\nabla \phi_{\alpha} + \vec{k}_{\alpha} + g\vec E + b \left(\vec{v}_{\alpha} \times \vec B\right).
\ee 

\ni If we compare the above equation with the standard momentum equation for a charged fluid immersed in an electromagnetic field \cite{Thompson}, we can see that the last two terms on the right side of Eq. (\ref{54}) is the Lorentz force. Thus the coupling constant $g$ and $b$ are both equal and given by 

\be\label{55}
g = b = {{\epsilon_{\alpha}}\over{m_{\alpha}}}
\ee

\ni where $\epsilon_{\alpha}$ is the charge and $m_{\alpha}$ is the mass of the charge. So, the Navier-Stokes equation of the fluid that can be written as             

\be\label{56}
{{\partial \vec{v}_{\alpha}}\over{\partial t}}+  \vec{\omega}_{\alpha} \times \vec{v}_{\alpha} = -\nabla \phi_{\alpha}  + {{\epsilon_{\alpha}}\over{m_{\alpha}}}\left[ \vec E +  \vec{v}_{\alpha} \times \vec B\right]+ \vec{k}_{\alpha}\,\,,
\ee 

\ni which is a very interesting result, where we have a Lagrangian density that describes a compressible and charged fluid in a very general representation. 

Using the vector and scalar potential ($\Phi$) of the electromagnetic field in (\ref{51}) we can rewrite the Lagrangian as follows

\be\label{57}
{\cal L}^{\prime} ={{1}\over{2}}\left[-{{\partial \vec{v}_{\alpha}}\over{\partial t}}-\nabla \phi_{\alpha} + \vec{k}_{\alpha} +{{\epsilon_{\alpha}}\over{m_{\alpha}}}\left( -{{\partial \vec A}\over{\partial t}}-\nabla \Phi \right) \right] ^2 -{{1}\over{2}}a_{\alpha}^2 \left(\nabla \times \vec{v}_{\alpha} +  {{\epsilon_{\alpha}}\over{m_{\alpha}}}\nabla \times \vec A \right)^2,
\ee

\ni which can be written as

\be\label{58}
{\cal L}^{\prime} ={{1}\over{2}}\left(-{{\partial \hat{\vec{v}}_{\alpha}}\over{\partial t}}-\nabla \hat{\phi}_{\alpha} + \vec{k}_{\alpha} \right)^2 -{{1}\over{2}}a_{\alpha}^2 \left(\nabla \times \hat{\vec{v}}_{\alpha}  \right)^2,
\ee

\ni where 

\be\label{59}
\hat{\vec v}_{\alpha} = \vec{v}_{\alpha} +{{\epsilon_{\alpha}}\over{m_{\alpha}}}\vec{A}
\,\,\,\,\qquad{\rm and}\,\,\,\,\qquad
\hat{\phi}_\alpha =\phi_\alpha +{{\epsilon_{\alpha}}\over{m_{\alpha}}}\Phi.
\ee

Notice that the Lagrangian in Eq. (\ref{58}) has the same structure as the Lagrangian in (\ref{35}) where the electromagnetic field is zero.

However, as we will see, the coupling introduces some modifications to the Maxwell-type equations of the charged fluid dynamics  shown in 
(\ref{44})-(\ref{47}). Using Eqs. (\ref{49}) and (\ref{50}) we have that

\be\label{60}
\hat{\vec l}_{\alpha} = -{{\partial \hat{\vec v}_{\alpha}}\over{\partial t}} -\nabla \hat{\phi}_{\alpha} +\vec{k}_\alpha
\ee

\ni and

\be\label{61}
\hat{\vec \omega}_{\alpha} =\nabla \times \hat{\vec{v}}_{\alpha}.
\ee

\ni Let us take the divergence, curl, and time derivative of the Lamb vector in Eq. (\ref{60}) and finally the divergence of the vorticity, we obtain

\be\label{62}
\nabla \cdot \hat{\vec l}_\alpha = \hat{N}_\alpha + {{\epsilon_{\alpha}}\over{m_{\alpha}}}{{1}\over{\epsilon_{0}}} \rho_{el}+\nabla \cdot \vec k_\alpha\,\,,
\ee

\ni and using both the Ampere law and

\be\label{63}
\hat{N}_\alpha = -{{\partial }\over{\partial t}}\nabla \cdot \hat{\vec v}_\alpha -\nabla^2 \hat{\phi}_\alpha ,
\ee

\ni we have that

\be\label{64}
\nabla \times \hat{\vec l}_\alpha +{{\partial \hat{\vec \omega}_\alpha}\over{\partial t}} =\nabla \times \vec{k}_\alpha.
\ee

\ni Substituting Eq. (\ref{64}) in the time derivative of the Lamb vector in the following way we can write that

\ba\label{65}
{{\partial \hat{\vec l}_\alpha}\over{\partial t}} &=&{{\partial \hat{\vec\omega}_\alpha}\over{\partial t}}\times \hat{\vec v}_{\alpha} +\hat{\vec\omega}_{\alpha} \times {{\partial \hat{\vec v}_\alpha}\over{\partial t}}\nonumber\\
&=&(\nabla \times \hat{\vec l}_\alpha\,) \times \hat{\vec v}_{\alpha} +\hat{\vec\omega}_{\alpha} \times {{\partial \hat{\vec v}_\alpha}\over{\partial t}},
\ea

\ni and after some calculation we can see that 

\be\label{66}
{{\partial \hat{\vec l}_\alpha}\over{\partial t}} -a^2_\alpha \nabla \times \hat{\vec\omega}_\alpha = - \hat{\vec{J}}_\alpha + {{\partial \vec{k}_\alpha}\over{\partial t}} ,
\ee

\ni where the new equation for the current density is given by

\be\label{67}
\hat{\vec J}_\alpha = \left(\hat{N}_\alpha + {{\epsilon_{\alpha}}\over{m_{\alpha}}}{{1}\over{\epsilon_{0}}} \rho_{el}+\nabla \cdot {\vec k}_\alpha\right)\hat{\vec v}_\alpha + 2\hat{\vec l}_\alpha \cdot \nabla \hat{\vec v}_\alpha -\hat{\vec l}_\alpha (\nabla \cdot \hat{\vec v}_\alpha) + \hat{\vec l}_\alpha \times \hat{\vec\omega}_\alpha - \hat{\vec\omega}_\alpha \times {{\partial \hat{\vec v}_\alpha}\over{\partial t}}\,\,.
\ee

\ni Finally, substituting the divergence of the vorticity in Eq. (\ref{61}) we obtain the last equation

\be\label{68}
\nabla \cdot \hat{\vec \omega}_{\alpha} = 0
\ee

\ni To sum up, we have a set of equations given by

\be\label{69}
\nabla \cdot \hat{\vec l}_\alpha = \hat{N}_\alpha + {{\epsilon_{\alpha}}\over{m_{\alpha}}}{{1}\over{\epsilon_{0}}} \rho_{el}+\nabla \cdot {\vec k}_\alpha,
\ee

\be\label{70}
\nabla \times \hat{\vec l}_\alpha +{{\partial \hat{\vec \omega}}\over{\partial t}} =\nabla \times \vec{k}_\alpha,
\ee

\be\label{71}
{{\partial \hat{\vec l}_\alpha}\over{\partial t}} -a^2_\alpha \nabla \times \hat{\vec\omega}_\alpha = - \hat{\vec{J}}_\alpha + {{\partial \vec{k}_\alpha}\over{\partial t}},
\ee

\be\label{72}
\nabla \cdot \hat{\vec \omega}_{\alpha} = 0
\ee

In the same way as in (\ref{37}), we can introduce a second-rank tensor taking into account the coupling with electromagnetic field, given by

\be\label{73}
G_{\alpha}^{\mu\nu} = T_\alpha^{\mu\nu} + {{\epsilon_\alpha}\over{m_\alpha}}F^{\mu\nu},
\ee

\ni where

\be\label{74}
T_\alpha^{\mu\nu} =\tilde{\partial}^\mu U_\alpha^\nu -\tilde{\partial}^\nu U_\alpha^\mu,
\ee

\ni and 

\be\label{75}
U_\alpha^{\mu\nu} =(\hat{\phi}_\alpha,\,\, a\hat{\vec{v}}_\alpha ),
\ee 

\ni where $F^{\mu\nu}$ is the second rank tensor of the electromagnetic field.

We can then rewrite the Lagrangian density (\ref{40}) as

\be
\label{76}
{\cal L}^{\prime} =-{{1}\over{4}}G_{\alpha}^{\mu\nu}G_{(\alpha)\mu\nu} -{{1}\over{a}} {J}_{\alpha}^{\mu}U_{(\alpha)\mu},
\ee

\ni where ${J}_{\alpha}^{\mu} = (\hat{N}_\alpha,\,\, a\hat{\vec J}_\alpha )$.

\section{Conclusions}

In \cite{Marmanis}, the author developed an analogous Maxwell-type formalism for an incompressible fluid without considering dissipation terms.   In this paper we believe that we have given a step forward in the issue since we have carried out a Maxwell-type electromagnetic model for a compressible fluid with dissipation terms from the very beginning of the calculations.

In order to present the formalism in a complete manner, we have also constructed the Lagrangian for this fluid and after that, for a charged one.   The ``configuration space" for both is formed by the Lamb vector and vorticity, which could be organized as components of a field strength tensor and in this way, the analogy with the Maxwell Lagrangian is direct.

Besides, in order to analyze the applications of this fluid ``electromagnetism" in QGP, we have accomplished a NA version of the uncharged compressible fluid.  The same procedure to the charged one can be obtained, as we have shown.

As an obvious perspective. we can use some analysis of the Yang-Mills theories to attack the QGP problem.   Besides, we can also consider other analogies for alternative electromagnetic formalisms like Podolsky, Lee-Wick and others.

\begin{acknowledgments}
\ni E.M.C.A. thanks CNPq (Conselho Nacional de Desenvolvimento Cient\' ifico e Tecnol\'ogico), a Brazilian research support agency, for partial financial support.
\end{acknowledgments}

\appendix*

\section{Non-Abelian fluids: a basic review}

The main motivation to investigate non-Abelian fluids is based on the increasing interest in the dynamics of NA plasmas at both very high temperature and density.  Other analysis can be carried out concerning relativistic NA plasmas in extreme cosmological and astrophysical conditions.   As examples, we can mention the electroweak plasma in the early Universe, the physics of the explosions occurring in supernovae and the physics of dense neutron stars.   So, it is important to have a complete theoretical description of a quantitative seceario for NA plasmas both in- and out- of equilirium  \cite{lm}.   Having said that, in this section we will promote a non-Abelianization of the main results obtained in the last section.
But, in the name of self-containement, let us review some few steps on non-Abelian fluids.

It is very well known that on deriving the equations of fluid mechanics from basic particle theory by statistical averages will apply just fine in the non-Abelian scenario \cite{jnpp}.   Considering, for example, the QGP, the one-particle kinectic equation can be written as \cite{jnpp}

\be
\label{AA}
P^{\mu}\, \bigglb[ \frac{\pa}{\pa ^{\mu}} \,+\, g\,Q_a F^a_{\mu\nu} \frac{\pa}{\pa P_{\nu}} \,+\, g f_{abc} A^b_{\mu} Q_c \frac{\pa}{\pa Q_c} \biggrb] f(X,P,Q)\,=\,C(t)
\ee

\ni where $f(X,P,Q)$ is the one-particle distribution function; $C$ is the collision integral term which considers the particle's scattering; $Aâ_{\mu}$ and $Fâ_{\mu\nu}$ are the potential and the field for a non-Abelian theory which rely on a gauge group with structure constants $F_{abc}$; and $Q_a$ is the classical colour charge of the particle.

The $C=0$ case means a collisonless plasma and the Boltzman equation (\ref{AA}) is the equation for the distribution function for single particles that obey the basic classical equations of motion for non-Abelian particles, namely, the Wong equation \cite{wong} given by

\be
\label{B.a}
\!\!\! m\,\frac{dX^{\mu}}{d\tau} \,=\,P^{\mu}\,\,,
\ee
\be
\label{B.b}
\qquad \quad m\,\frac{d P_{\mu}}{d\tau}\,=\,g\,Q_a F^a_{\mu\nu} P^{\nu}\,\,,
\ee
\be
\label{B.c}
\qquad \qquad \quad m\frac{d Q_a}{d\tau}\,=\,-\,g f_{abc} P^{\mu} A^b_{\mu} Q^c\,\,,
\ee

\ni where the motion of the color degrees of freedom is in a phase space way as the color index runs from $a=1$ to $N^2 -1$ for a $SU(N)$ gauge group and $\tau$ is the proper time of the particle.  We can say that the proper space is the Lie group modulo the maximal torus \cite{jnpp}.  Considering a microscopic analysis, the trajectories in phase space are known exactly.   The Wong equations give the trajectories $x(\tau),\,p(\tau)$ and $Q(\tau)$ for every particle, i.e., they are the classical equations of motion.   From (\ref{B.c}) \cite{lm} we can see that the NA charges are also subjected to the dynamical evolution.  The Eq. (\ref{B.c}) can be rewritten as $D_{\tau} Q_a =0$, where 
$D_{\tau}=\frac{d x^{\mu}}{d \tau}\,D_{\mu}$ is the covariant derivative in the world line.  And 
$$D_{\mu}^{ac}[A] = \pa_{\mu}\,\delta^{ac}\,+\,g\,f^{abc}\,A_{\mu}^b$$ is its adjoint representation \cite{lm}.   Besides, the Boltzmann equation (\ref{AA}) is invariant under gauge transformations, i.e., if $f(X,P,Q)$ is the solution of (\ref{AA}), $f(X,P,U^{-1}QU)$ is also a solution.

The fact that the equations for the Abelian fluid have a very large regime of validity, one can consider a derivation of a NA fluid mechanics, which encompasses the NA degrees of freedom, coupling to a NA gauge field, etc., which may be correct for dense, nonperturbative and nondilute systems.  The field strength $F^a_{\mu\nu}$ and the energy-momentum tensor of the gauge fields are written respectively as

\ba
\label{C}
F^a_{\mu\nu} [A] \,&=& \pa_{\mu} A^a_{\nu}\,-\,\pa_{\nu} A^a_{\mu}\,+\,g f^{abc} A^b_{\mu} A^c_{\nu} \nonumber \\
\Theta^{\mu\nu} [A] &=& \frac 14 g^{\mu\nu} F^a_{\rho\sigma} F^{\rho\sigma}_a \,+\, F^{\mu\rho}_a F^{a\nu}_{\rho}
\ea

\ni where $f^{abc}$ are the structure constants of $SU(N)$ and we are working in natural units.   The anti-symmetric structure constants $f_{abc}$ results from the commutator $[\lambda_a,\,\lambda_b ] = 2if_{abc} \lambda_c$, where $\lambda_a$ are the Gell-Mann matrices.   From (\ref{B.c}) we can see that the NA charges have also a dynamical evolution.  With the solutions of Wong's equations we can construct the color current \cite{lm} for the particles given by

\ba
\label{D}
j^{\mu}_a (x) &=&\, g \int d\tau \frac{d x^{\mu}}{d\tau} Q_a (\tau) \delta^{(4)} [x\,-\,\bar{x} (\tau) ] \nonumber \\
&=&g \int d\tau \frac{p^{\mu}}{m} Q_a (\tau) \delta^{(4)} [ x\,-\, \bar{x} (\tau) ]
\ea

\ni and we can use (\ref{B.a}) and (\ref{B.b}) to write

\be
\label{E}
m\ddot{x}^{\mu} (\tau)\,=\, g Q^a F_a^{\mu\nu} (x(\tau)) \dot{x}_{\nu} (\tau)
\ee

\ni where dots denote derivatives with respect to $\tau$.   Using the Wong equations we can find that $D_{\mu}\,j^{\mu}=0$, namely, the current is conserved.   The particles energy-momentum tensor for the particles is given by

\be
\label{F}
T^{\mu\nu}_{part.} (x)\,=\, \int d\tau \frac{dx^{\mu}}{d\tau} p^{\nu} (\tau) \delta^{(4)} [x\,-\,\bar{x} (\tau0 ]
\ee

\ni and the Yang-Mills equations are

\be
\label{G}
D_{\mu} F^{\mu\nu} (x) \,=\, J^{\nu} (x)
\ee

\ni which has the source term like

\be
\label{H}
J^{\nu} (x) \,=\, \sum_{particles} j^{\nu} (x)
\ee

\ni for the sum of all particles.   The Yang-Mills (color) field energy-momentum tensor is given by

\be
\label{I}
T^{\mu\nu}_{YM}\,=\, F_a^{\mu\rho} F^{a\nu}_{\rho} \,+\, \frac 14 g^{\mu\nu} F^2 \,\,,
\ee

\ni where $F^2 = F^{\mu\nu a}\,F^a_{\mu\nu}$.   Hence, we have the conservation law

\be
\label{J}
\pa_{\mu} [T^{\mu\nu}_{part.} (x)\,+\,T^{\mu\nu}_{YM} (x) ] \,=\,0
\ee

\ni where the divergence $\pa_{\mu}\,T^{\mu\nu}_{YM}$ is given by

\be
\label{K}
\pa_{\mu} T^{\mu\nu}_{YM}\,=\,g j^a_{\mu} F^{\mu\nu}_a
\ee

\ni where we have used that

\be
\label{L}
\Big(D_{\mu} F^{\mu\nu}\Big)_a (x) \,=\,\pa_{\mu} F^{\mu\nu}_a (x)\,+\, g\,f_{abc} A^b_{\mu}(x) F^{\mu\nu c} (x) \,=\, g\,j^{\nu}_a (x)
\ee

\ni where $j^{\mu}_a (x)$ is the color current for the particles given by Eq. (\ref{D}).   So, based on Eqs. (\ref{D}) and (\ref{J}) we can write that

\be
\label{M}
\int d\tau \frac{d x^{\mu}}{d\tau}\,Q_a (\tau) \delta^{(4)} [x-\bar{x}(\tau)]\,=\,0
\ee

\ni which is specifically true only in the classical field approximation to the color sector.   The conservation laws in Eqs. (\ref{J}) and (\ref{M}) can be used as standard for a macroscopic fluid dynamical analysis for the quark-gluon plasma.   However, it is important to talk about space-time flow of macroscopic variables (energy and momentum density), entropy density, temperature, etc. \cite{heinz2} which rules the general structure of the plasma lifetime and evolution.   In the set of these conservations laws we can notice that it shows the coupling of the colored quark-fluid to the NA color field, which is different from standard fluid dynamics, it is the so-called chromohydrodynamics, which is the NA extension of plasma physics.



\begin{thebibliography} {99}

\bibitem{jackiw}   R. Jackiw, ``Inserting Group Variables into Fluid Mechanics," arxiv: hep-th/0410284.

\bibitem{heinz}    U. Heinz, Phys. Rev. Lett. 51 (1983) 351.

\bibitem{choquet}   Y. Choquet-Bruhat, J. Kath. Phys. 33 (1992) 1782.

\bibitem{hk}   D. D. Holm and B. A. Kupershmidt, Phys. Rev. D 30 (1984) 2557.

\bibitem{patten}   M. H. P. M. van Putten, ``The Theory Of Ideal Yang-Mills Fluids In Symmetric Hyperbolic Form," arxiv: hep-ph/9310315.

\bibitem{bi}   J-P. Blaizot and E. Iancu, Nucl. Phys. B 421 (1994) 565.

\bibitem{Kambe}    T. Kambe, Fluid Dyn. Res. {\bf 42}, 055502 (2010).

\bibitem{Thompson}     R. J. Thompson and T. M. Moeller, Phys. of Plasmas {\bf 19}, 010702 (2012); {\bf 19}, 082116 (2012).

\bibitem{Marmanis}    H . Marmanis, Phys. Fluids. {\bf 10}, 1428 (1998).

\bibitem{lighthill}     M. J. Lighthill, Proc. R. Soc. Lond. A 211 (1952) 564; ibid 222 (1954) 1.





\bibitem{Landau}    L. D. Landau and E.M. Lifshits, {\it Fluid Mechanics} (Pergamon Press, Oxford, 1980).



\bibitem{Russakoff}    G. Russakoff, Am. J. Phys. {\bf 38}. 1188 (1970). 



\bibitem{Albert1}    A. C. R. Mendes, C. Neves. W. Oliveira and F.I. Takakura, Braz. J. Phys. {\bf 33}, 346 (2003).

\bibitem{Arnold}   V. I. Arnold, {\it Mathematical Methods of Classical Mechanics} (Springer, New York, 1989). 

\bibitem{arti18e19dobrazilian}    V. I. Arnold and B. A. Khesin, Ann. Rev. Fluid Mech. {\bf 24}, 145 (1992); V. Zeitlin, J. Phys. {\bf A24}, L171 (1992).


\bibitem{Mahajan}   S. M. Mahajan, Phys. Rev. Lett. {\bf 90}, 035001 (2003).

\bibitem{truesdell}    C. Truesdell, ``The kinematics of Vorticity," Indiana University Press, USA, 1954.

\bibitem{Curtis}    C. W. Hamman, J. C. Klewicki and R.M. Kirby, J. Fluid Mech. {\bf 610}, 261 (2008).

\bibitem{Xin}     X. Zhao and G-W. He, Phys. Rev. E {\bf 79}, 046316 (2009).

\bibitem{Frish}U. Frish, {\it Turbulence: The Legacy A.N. Kolmogorov} (Cambridge University Press, Cambridge, 1995).

\bibitem{jnpp}   R. Jackiw, V. P. Nair, S. Y. Pi and A. P. Plolychronakus, J. Phys. A 37 (2004) R327.

\bibitem{wong}    S. K. Wong, Nuovo Cimento A 65 (1970) 689.

\bibitem{lm}    D. F. Litim and C. Manuel, Phys. Rept. 364 (2002) 451.

\bibitem{heinz2}    U. Heinz, Ann. Phys. 161 (1985) 48.

\bibitem{bmm}   B. A. Bambah, S. M. Mahajan and C. Mukku, Phys. Rev. Lett. 97 (2006) 072301.

\end{thebibliography}
\end{document}